\RequirePackage[2020-02-02]{latexrelease}
\documentclass[showpacs,preprintnumbers,amsmath,amssymb]{revtex4}
\newcommand{\ket}[1]{\left | #1 \right\rangle}
\newcommand{\bra}[1]{\left \langle #1 \right |}

\usepackage{dcolumn}
\usepackage{bm}
\usepackage[dvips]{graphics}
\usepackage{tikz}
\usetikzlibrary{quantikz}
\usepackage{natbib}

\begin{document}

\title{Observing ghost entanglement beyond scattering amplitudes in quantum electrodynamics}

\author{Chiara Marletto and Vlatko Vedral}
\affiliation{Clarendon Laboratory, University of Oxford, Parks Road, Oxford OX1 3PU, United Kingdom}

\date{\today}

\begin{abstract}
A fully local quantum account of the interactions experienced between charges requires us to use all the four modes of the electromagnetic vector potential, in the Lorenz gauge. However, it is frequently stated that only the two transverse modes of the vector potential are ``real" in that they contain photons that can actually be detected. The photons present in the other two modes, the scalar and the longitudinal, are considered unobservable, and are referred to as ``virtual particles" or ``ghosts". Here we argue that this view is erroneous and that even these modes can, in fact, be observed. We present an experiment which is designed to measure the entanglement generated between a charge and the scalar modes. This entanglement is a direct function of the number of photons present in the scalar field. Our conclusion therefore is that the scalar quantum variables are as ``real" as the transverse ones, where reality is defined by their ability to affect the charge. A striking consequence of this is that we cannot detect by local means a superposition of a charge bigger than that containing 137 electrons.
\end{abstract}

\pacs{03.67.Mn, 03.65.Ud}

\maketitle

Known quantisation procedures usually start with a classical theory and proceed to its quantum version by promoting relevant physical quantities to operators obeying a non-commutative algebra. In electromagnetism, in order to maintain locality of interactions, it is necessary to carry out quantisation by using the electromagnetic (EM) potentials, expressed in a gauge that can be chosen according to various criteria. The Lorenz gauge, in particular, is representative of a class of gauges that is explicitly Lorentz-covariant. The price to pay for Lorentz-covariance is the appearance of the so-called ``ghost modes" (one longitudinal and one scalar): they are called `ghosts' because they are considered unobservable dynamical variables, that need to be removed by means of imposing supplementary conditions (the so-called `Lorenz gauge conditions) to recover the space of physically relevant observables, which in the case of electromagnetism are the transverse modes.  What decides whether given modes are physically relevant or not is whether they are gauge-invariant. Transverse modes of the potentials are the only gauge-invariant modes, which are present in all gauges and hence are traditionally used to define the space of physical observable states. Some have expressed dissatisfaction with the existence of ghost modes, as they appear to be redundant; while others have argued that they are physical and they are necessary to have a fully local, relativistically compliant account of electromagnetism with sources. 

In this paper we uncover a potentially groundbreaking fact, which suggests that ghost modes are not physically irrelevant as usually thought. We shall explain that in the Coulomb interaction of a quantised charge with the EM potentials expressed in the Lorenz gauge, the charge and the EM field are entangled, and that this entanglement is, in fact, observable. Specifically, we suggest an interference experiment involving a charge superposed across two locations, which uses quantum state tomography to detect this entanglement. This experiment amounts to (indirectly) detecting the scalar ghost modes. We say that the detection is `indirect' because it does not correspond to detecting single photons associated with such modes. Nonetheless, this entanglement is an (at least in principle) observable effect; and while it is predicted in the Lorenz gauge, it is not predicted in other gauges (such as the Coulomb gauge) as traditionally quantised, where such modes do not exist. This is so unless, of course, one acknowledges the quantum nature of all the 4 modes in all gauges, but then proceeds to define different constraints that would correspond to different gauges. We note the spirit of our conclusion is similar to other analysis about ghost modes emerging in other gauges, such as the longitudinal modes in the temporal gauge, \cite{KAY1, KAY2}. 

This important effect has so far been unnoticed because, to predict it, one needs to go beyond standard procedures such as the S-matrix methods. The quantity that we propose to measure is not a scattering amplitude as traditionally conceived in quantum field theory, (see e.g. \cite{WEI}). This is because it uses an {\sl open-loop} interferometry on the charge, which, as we shall explain, is different from a traditional closed loop interferometry. Therefore, our proposed experiment cannot be described in terms of the free input and output fields as is normally done in quantum field theory. 

This result leaves us with a few possible alternatives: 1) physical ghost modes (such as the scalar modes in the Lorenz gauge, which propagate causally) are in fact observable and only predicted in some gauges; hence gauge-invariance must be understood as valid only in some classical limit, but violated in the quantum case;  2) all modes must be quantized in all gauges; however, the price to pay is that the resulting entanglement may not satisfy the standard observability requirements (to be be clarified later); 3) the entanglement with the ghost modes is not observable even in principle; 4) the quantisation scheme for the EM field as currently known is invalid (as it breaks gauge invariance), so one needs to look for alternative ways to construct a quantum theory of fields. In the discussion section we consider these alternatives and we also consider the implications of these observations for the recently-proposed witnesses of non-classicality based on entanglement generation via a mediator. 

{\bf Ghost entanglement: informal description.} In any gauge, one can always represent the quantised modes of the free field as represented by a set of simple quantum harmonic oscillators, one per each degree of freedom (or mode). The $x$ and $p$ variables are simply the quadratures of the field, and in the ground state, the dispersion in $x$ is proportional to $\Delta x \propto 1/\sqrt{\omega}$, where $\omega$ is the frequency of the mode. When a charge is introduced, each of the modes responds because the presence of the charge acts as a perturbation. 

In the static regime, the scalar modes of the electromagnetic field as described in the Lorenz gauge, become a collection of driven harmonic oscillators, with the driving term equal to $-qx$. This results in the ground state of the oscillators being displaced to a coherent state. The displacement in the position becomes $\delta x \propto 1/\omega^2$. The amplitude of this coherent state is $\lambda_\omega\propto \Delta x /\delta x\propto 1/\omega^{3/2}$ and therefore it has $n_\omega =|\lambda_\omega |^2 \propto 1/\omega^{3}$ photons on average. This is the probability that the new ground state of the driven field mode will result in the excited state of the free field (without interactions). 

These states of scalar photons are considered virtual and undetectable because they do not contain any transverse photons; nonetheless, they are necessary to give a local, causal and Lorentz-covariant account of the Coulomb interaction between static sources. Following this account, when a charge is in a superposition of two different positions,  each of the positions generates its own set of coherent states (whose amplitude peaks at the location of the charge). In other words, the charge and the scalar modes of the field become (at least formally) entangled. The amount of entanglement can be computed exactly (as we will show below) and it is equal to the total number of scalar photons $n=\alpha \int |\lambda_\omega |^2 d^3 \omega = \alpha \int d \omega/\omega$,
where $\alpha$ is the fine structure constant. This integral is logarithmically divergent (the notorious ``infrared catastrophe"), however, with reasonable cut-offs it can be argued that it is on the order of unity (with logarithmic corrections as a function of the extent of the superposition). The amount of entanglement between the charge and the field is therefore very closely approximated by the fine structure constant. It is for this reason that, when the charge is bigger than the inverse of the fine structure constant times the electron charge, the entanglement with the scalar modes becomes close unity thereby significantly decohering the superposition of the charge. 

Hereinafter we shall treat the charges non-relativistically. This is because they are considered to be (nearly) stationary. Formally it also means that there will be an energy cut-off of $m_ec^2$ and that no field modes within the radius equal to the Compton wavelength of the electron will be included in the analysis. This is not an important omission since the relativistic effects neglected here would represent a small correction to the value we estimated.   

The crucial question now is: can this field-matter entanglement (as formally predicted in the Lorenz gauge) be detected? Naively, one might think that this entanglement cannot be detected. First of all, the photons are associated with the `ghost' modes (the scalar modes in the Lorenz gauge) and should therefore not be capable of generating `clicks' in photodetectors (only the excitations of the transverse modes make clicks). In other words, the scalar modes do not contain free photons: in a dynamical situation scalar and longitudinal photons always cancel out, which is (as we shall discuss) the condition defining the Lorenz gauge. 
Secondly, the entanglement between the charge and the field should lead to what is known as the `fake' (false) decoherence. This means that, although normally entanglement between a system (in this case the charge) and its environment (in this case the scalar modes) should lead to a decrease of the interference fringes (and hence decoherence), here, when the charge is interfered, the two sets of coherent states (each generated in the two branches of the charge's quantum state) become one and the same state because the two spatial states of the charge are rejoined at the point of interference. Therefore, in a single charge (and `closed loop') interference experiment, the charge entanglement with the scalar modes is simply undetectable since it does not affect the (closed loop) interference. It is our opinion that all the charge interference experiments so far have always been of this kind, which is why the scalar mode entanglement remained unnoticed. 

The above arguments are sound, and yet we believe that the field-matter entanglement can be detected. This is because the arguments apply to what one may call a 'closed loop' interference on the charge. It is possible, however, to measure the charge-field entanglement {\sl without closing} the interferometric loop of the charge, using a standard technique of quantum metrology, called quantum state tomography. We discussed this experimental setup in our analysis of the locality of the Aharonov-Bohm effect, \cite{MV-AB}. Specifically, we need two charges (one being a reference, the other the probe), each superposed across two locations; and we use this method to perform a tomographic reconstruction of the probe's state using local measurements on each path, and the second charge as a reference. The quantum state of the probe is predicted to be mixed, due to the entanglement with the field. Hence this is a way to indirectly confirming entanglement between the probe charge and the field (assuming no other decoherence is at play). 

For convenience, let us divide the space where the interference takes place into two regions, labelled `left' and `right', with respect to the line that cuts the interferometer into two symmetric halves (see the figure). 

\begin{figure}[h]
\includegraphics[width=8cm]{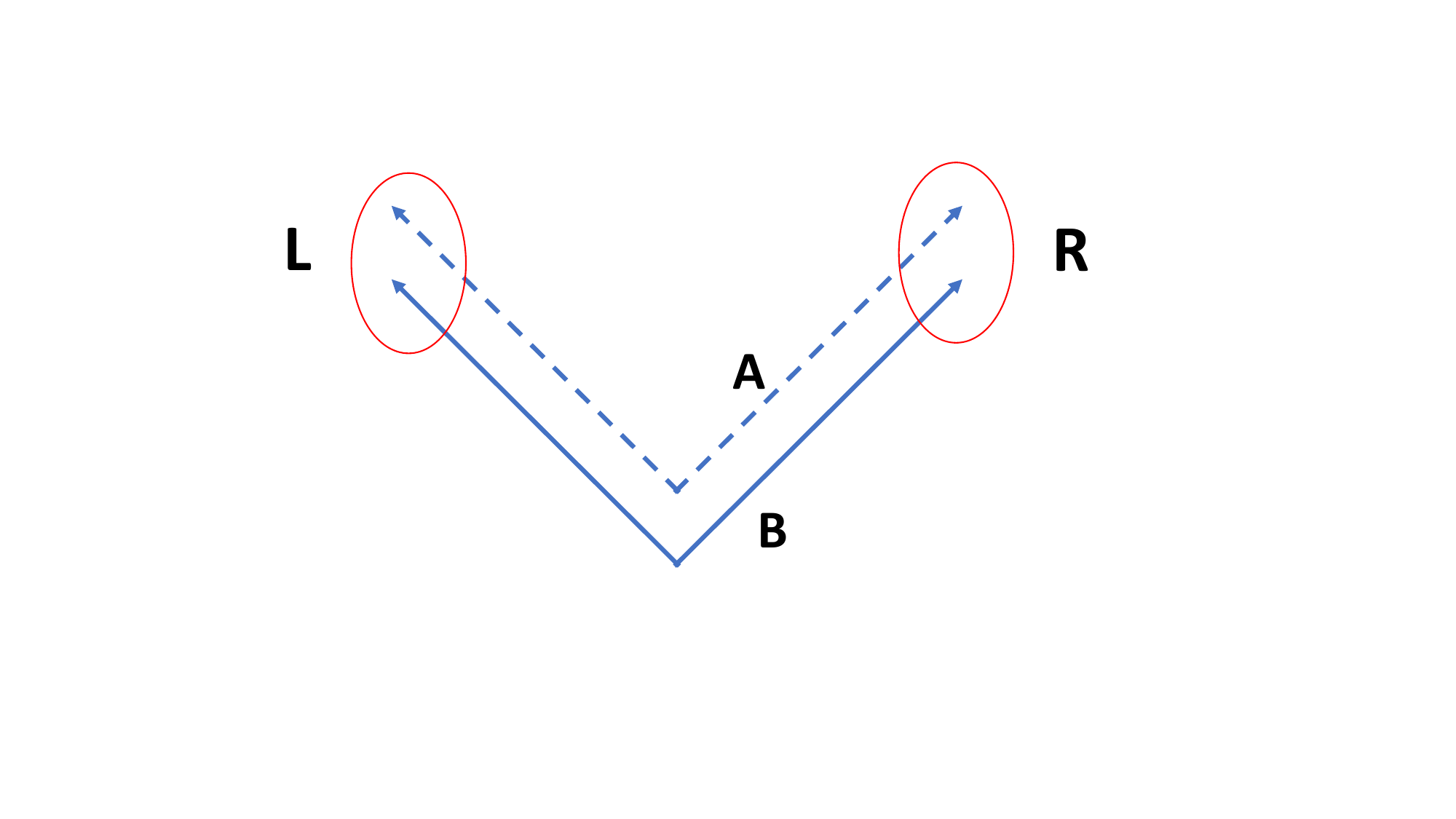}
\caption{Both charge A and charge B (the reference) are superposed across regions L and R. Joint measurements of A and B, local to the left and right regions, reconstruct the Coulomb phase as well as the reduced visibility due to entanglement of charge A and B with the scalar modes.}
\label{PiC}
\end{figure}
Suppose that the charge A (probe) is at locations $AR$ and $AL$ and the second one (reference) is at locations $BR$ and $BL$. We then have four spatial configurations of charges, $|AR,BL\rangle, |AR,BR\rangle, |AL, BL\rangle, |AL,BR\rangle$, each entangled to a different corresponding set of scalar modes $|\lambda_{AR,BL}\rangle, |\lambda_{AR,BR}\rangle, |\lambda_{AL, BL}\rangle, |\lambda_{AL,BR}\rangle$. 

By making suitable local measurements with respect to the partition Left/Right, on both charges, we can obtain the off-diagonal element in the subspace where there is only one charge in modes $AR$ and $AL$ and only one charge in modes $BR$ and $BL$. This element contains a pure phase, which is due to the Coulomb interaction, and a real amplitude which represents the entanglement with the scalar modes. It is this real amplitude that is equal to $\langle \lambda_{AR, BL}|\lambda_{AL, BR}\rangle$ and effectively shows that the scalar photons should be considered as ``real" as the transverse ones (although they do not propagate at the speed of light and are always linked to the charge). We proceed to discuss the details of this scheme in the remaining part of the paper, while in the appendix we have summarised the standard quantisation in the Lorenz gauge, as described in textbooks (e.g. \cite{COH}).

{\bf Measuring ghost entanglement with an open loop tomography.} Let $b_{AL}^{\dagger}$, $b_{AL}$  be bosonic creation/annihilation operators for the charge $A$ to be in a spatial mode $r_L$ on the left; and $b_{AR}^{\dagger}$, $b_{AR}$ be bosonic creation/annihilation operators for the charge $A$ to be in a spatial mode $r_R$ on the right; and similarly for charge $B$. For instance, the state where the charge A is superposed across two locations is $\frac{1}{\sqrt{2}}(b_{AL}^{\dagger}+b_{AR}^{\dagger})\ket{0}$ where $\ket{0}$ is the bosonic vacuum. 
When the charge $A$ is prepared in a superposition of the left and right path, charge-field Hamiltonian will induce the state: $\frac{1}{\sqrt{2}}(b_{AL}^{\dagger}\ket{0}\ket{\lambda_{AL}}+\exp(i\Delta\phi)b_{AR}^{\dagger}\ket{0}\ket{\lambda_{AR}})$ where the Coulomb phase $\Delta \phi$ is a function of the points ${\bf r_L},{\bf r_R}$ across which the charge is superposed. 

Measuring the degree of entanglement between the charge $A$ and the field directly by local tomography on the charge $A$ is impossible, because of charge superselection rules, which impede measurements of observables such as $b_{AL}+b_{AL}^{\dagger}$. However, as pointed out in \cite{MAVEAB}, the degree of entanglement can still be reconstructed by utilising another reference charge, and local tomography (on the left and right sides) involving the same number of charges, without violating any superselection rule. This is an effective way of measuring the phase {\sl without closing the interferometer loop coherently} - i.e., with only decoherent communication between the two sides. 

To implement this procedure, we follow the protocol described in \cite{MAVE}. Consider the state where two charges, $A$ and $B$, are each superposed across the left and right regions:

\begin{eqnarray}
&\frac{1}{{2}}\times& (\exp(i \phi_{R,R})b_{AR}^{\dagger}b_{BR}^{\dagger}\ket{0}\ket{\lambda_{AR, BR}}+ \exp(i \phi_{R,L})b_{AR}^{\dagger}b_{BL}^{\dagger}\ket{0}_A\ket{\lambda_{AR, BL}}\nonumber \\
&+&
\exp(i \phi_{L,R})b_{AL}^{\dagger}b_{BR}^{\dagger}\ket{0}\ket{\lambda_{AL, BR}}+\exp(i \phi_{L,L})b_{AL}^{\dagger}b_{BL}^{\dagger}\ket{0}\ket{\lambda_{AL,BL}})
\end{eqnarray}

where $\exp(i\phi_{i,j})$ is the Coulomb phase relative to the configuration of charges $i,j$ as computed in the previous section. One can group the terms relative to the left and to the right modes, as follows:

\begin{eqnarray}
&\frac{1}{{2}}\times& (\exp(i \phi_{R,R})b_{AR}^{\dagger}b_{BR}^{\dagger}\ket{0}\ket{\lambda_{AR, BR}} +  \exp(i \phi_{L,L})b_{AL}^{\dagger}b_{BL}^{\dagger}\ket{0}\ket{\lambda_{AL,BL}}\nonumber \\
&+&\exp(i \phi_{R,L})b_{AR}^{\dagger}b_{BL}^{\dagger}\ket{0}_A\ket{\lambda_{AR, BL}}+\exp(i \phi_{L,R})b_{AL}^{\dagger}b_{BR}^{\dagger}\ket{0}\ket{\lambda_{AL, BR}})\;.\nonumber \\
\end{eqnarray}

In the branches where only one charge is present on the left and right arms (second line of the above equation) the degree of entanglement with the field can be detected by measuring, locally on the left and on the right, the $\hat C_{RL}=b_{BL}^{\dagger}b_{AL} b_{BR}b_{AR}^{\dagger} + b_{BL}b_{AL}^{\dagger} b_{BR}^{\dagger}b_{AR}$: if there were no entanglement with the field, for the above state this observable would be sharp with value $1$ in the subspace where only one charge is on the left and one on the right. Due to the entanglement with the EM field, as prescribed by the Lorenz gauge condition \eqref{HS}, the expected value of this observable is not $1$, but it is attenuated by the overlap of the coherent states of the field in that subspace, $\langle \lambda_{AR,BL}|\lambda_{AL,BR}\rangle$, whose value can be calculated formally from the definitions given in the appendix. For generic locations $r_a$ and $r_b$, one has

\begin{equation}
\langle{\lambda_{a}}|{\lambda_{b}}\rangle= \exp(-\frac{1}{2}|\lambda_{a}-\lambda_{b}|^2) =\frac{2g(k)^2}{(\hbar\omega(k))^2 }\left( 1-\cos \left (  \vec k \cdot (\vec r_a-\vec r_b)\right)\right )\;.
\end{equation}

By integrating in $d^3k$, we see that this is a (logarithmically) divergent integral. It is therefore meaningful to introduce a lower and an upper cutoff, in which case we find that the integral is proportional to the fine structure constant times $\ln (\Delta r/r_0)$ where $r_0$ is some minimum spatial extension (corresponding to the upper momentum cutoff) and $\Delta r = |\vec r_a-\vec r_b|$ is the extent of the superposition (corresponding to the lower momentum cutoff).  We thus see that the dependance on the extent of the superposition is only logarithmic, and therefore dominated by the fine structure constant. It seems meaningful to assume that $r_0=\hbar/mc$ is given by the Compton wavelength of the electron, since, as we explained, our treatment of the electron is non-relativistic.  But, because of the logarithmic dependence, this choice will anyway not affect the outcome a great deal (another possible choice is, of course, the size of the spatial resolution of the detector). 

Measuring $\hat C_{RL}$ by local measurements does not violate charge conservation superselection rule, as it can be done by resorting to local observables (with respect to the partition Left/Right) that include the same charge number. By local tomographic reconstruction of the 1-particle sector of the above state, one can therefore notice this reduction in visibility at any point along the path, without closing the interferometer coherently. Note, again, that it is crucial not to close the loop of the interferometer, because if one did that, the entanglement would vanish due to the fake decoherence effect. 

Our conclusions would be applicable to the linearised quantum gravity in the corresponding gauge to Lorenz, where the calculation is more complicated but the logic is essentially the same. The difference would be the magnitude of the entanglement, which would be present, but (unlike in this case) more difficult to observe. In fact, we previously showed that the gravitational entanglement with a mass $m$ scales as $1-e^{(m/m_P)^2}$, where $m_P$ is Planck's mass. It is, therefore, natural to suppose that the electromagnetic entanglement with the charge will scale as $1-e^{(Q/Q_P)^2}$, where $Q_P$ is the Planck charge. If $Q=e$, then $(e/Q_P)^2=\alpha$, which is our result to the lowest order. 

So, the same way that the entanglement with the field would prevent us from observing locally a charge superpostion beyond a certain limit, the same would be true gravitationally, where the limit would be then given by the Planck mass of the object. 

{\bf Discussion.} We have demonstrated that, by following standard quantisation procedures for the EM field, in the Lorenz gauge, the scalar modes of the EM field are entangled with any superposed charge, and we have also discussed a possible experiment to detect this entanglement. It is interesting that in the derivation we discussed the entanglement is induced by the Lorenz condition \eqref{HS} in the presence of static sources and by the possibility of superposing charges across different spatial regions. 

Broadly speaking, as we said in the introduction, this result can be interpreted in four different ways. 

One is that gauge invariance is broken. In the Coulomb gauge, at least in the case of straightforward quantisation procedures, this entanglement cannot exist, because the scalar and longitudinal modes are not quantised. In the temporal gauge one can find entanglement, but it is with the longitudinal modes of the vector potential \cite{Jackson, KAY1}. These modes are considered not physical either (because they propagate at a speed that is lower than the speed of light), but they would appear to generate a similar entanglement to that due to the scalar modes in the Lorenz gauge. The fact that different gauges predict different degrees of entanglement implies that the quantised theory of the EM field violates gauge invariance, as detecting this entanglement would allow us to discriminate between different gauges. These considerations therefore suggest that the current quantised theory of the EM field violates gauge-invariance, which must be only valid in the classical limit. This may imply (as some authors have argued) that the photon mass is non-zero, and that it could be related to the degree of entanglement with the charge.

One way out of this (and this is option 2 in the introduction) is of course to always quantize either the longitudinal and/or the scalar modes in all gauges. For instance, in the coulomb gauge the scalar potential is instantaneous, but can be written as an operator $A_0 (y) =\int dx \rho (y)/|x-y|$. This is why, as we said above, entanglement is also predicted in the temporal gauge, since there the longitudinal component $A_L$, is a quantum mechanical operator. As is well known, even though some potentials in gauges other than Lorenz violate microcausality, this is not a problem because all the observable physical effects always rely on all the components in the form $j_\mu A^\mu$. 

Another possible scenario (option 3) is that it is incorrect to consider the charge and the field as independent subsystems. In other words, while formally they may be described by commuting operators in the current quantum theory of light, charges and fields do not exist independently of one another, hence the entanglement in question may be unobservable after all. According to this view, real charges would be the bare (mathematical) charges dressed with the scalar photons. Neither the bare charges, nor the scalar photons can ever exist on their own. In this sense, the operator we claimed we were measuring in our open loop experiment would already contain dressed operators. The whole procedure of dressing is akin to mass renormalisation and it would lead us to conclude that the Coulomb gauge and the Lorentz gauge are in agreement after all because here no gauge predicts any entanglement.    

Finally (option 4) it could also be that the quantised theory we are using is not the correct way to design a quantum theory of the electromagnetic field. This is an intriguing possibility, and it is supported by the fact that even the founding fathers of QED considered it an effective theory that one could use to perform calculations, to be then replaced by a full quantum theory of fields. What this theory might be is an open problem. One could speculate that only once a full quantum theory of spacetime is available, one will be able to understand quantum fields on quantum spacetime (such as the EM field) properly, and that those quantum theories will differ radically from the quantised theories we currently have.

These open questions will, as always, only be settled by a combination of new theory and novel experiments. The experiment we discussed is a first step to assessing the validity of some of these conjectures, and it is within direct reach of current technologies: we hope and expect it will be performed as soon as possible in the near future.

\textit{Acknowledgments}: The Authors are grateful to David Deutsch and Bernard Kay for illuminating discussions. CM thanks the Eutopia foundation. This research is funded in part by the Gordon and Betty Moore Foundation. This publication was also made possible through the support of the ID 61466 grant from the John Templeton Foundation, as part of the The Quantum Information Structure of Spacetime (QISS) Project (qiss.fr). The opinions expressed in this publication are those of the Authors and do not necessarily reflect the views of the John Templeton Foundation.

\section{Appendix: Formal calculation of the entanglement in the Lorenz gauge} 

In this appendix we summarise standard material that appears in quantum electrodynamics textbooks, such as \cite{COH}, and we apply it to the calculation of the field matter entanglement as presented in the main text. For convenience, we follow the presentation given in \cite{MAVE23}. 

We denote by ${\vec r_i}$ a particular 3-vector representing the position of a charge. We shall also denote by $b_i$, $b_i^{\dagger}$ the bosonic annihilation and creation operators relative to the charge located at ${\vec r_i}$. The charge density operator for a two-charge distribution reads:

\begin{equation}
\hat \rho ({\vec r_{ij}}) = q \left (b_i^{\dagger}b_i\delta({ \vec r}-{ \vec r_i})+b_j^{\dagger}b_j\delta ({ \vec r}-{ \vec r_j}) \right )\;.
\end{equation}

One of the charges is the reference and the other is the probe, as described in the main text. We shall use a general charge $q$, but for the main text we have specialised to the case where $q$ is the electron.  
Its Fourier transform in momentum space (as a function of the 3-vector $\vec k$ representing the momentum) is:

\begin{equation}
\hat\rho_{ij} (\vec {k}) = \frac{1}{{(2\pi)}^{\frac{3}{2}}}q (b_i^{\dagger}b_i\exp{(-i{ \vec k\cdot \vec r_i})}+b_j^{\dagger}b_j\exp{(-i{ \vec k \cdot\vec r_j})})\;.
\end{equation}

The total Hamiltonian in the Lorenz gauge is $$H_{L}=H_F+H_I\;,$$ where $H_F$ is the free Hamiltonian of the field and $H_I$ is the interaction Hamiltonian. 

The free Hamiltonian $H_F$ has the form:

\begin{equation}
H_F= \int {\rm d}^3k\hbar \omega(k) \left(\sum_{\nu=1}^{3}a_{\nu}(\vec k)^{T}a_{\nu}(\vec k) -  a_{0}(\vec k)^{T}a_{0}(\vec k)\right)\;.
\end{equation}

Here, $a_{\mu}(\vec k)$ is the annihilation operator for the k-th mode of $\mu$ component of the electromagnetic field, where $\mu=0, 1, 2, 3$, while $a_{\mu}(\vec k)^{T}$ is its adjoint; and we have defined $\omega(k)=ck$, ($k= |\vec k|$). 

Assuming the metric tensor to be that of a flat Minkowski spacetime implies that the creation and annihilation operators for the scalar modes must obey the modified commutation relation: $$[a_{0}(\vec k), a_{0}(\vec k)^{T}]=-1\;.$$ 

As a consequence, the scalar odd number states $\left (a_{0}(\vec k)^{T}\right)^{2n+1}\ket{0}$ have a negative norm if the norm is defined as usual. This problematic issue is fixed by imposing a supplementary condition, ensuring that Gauss' law is satisfied at all times during the unitary dynamical evolution by simply annihilating the states that do not satisfy it. For the free field, the condition is the quantum equivalent of the classical constraint defining the Lorenz gauge, $\sum_{\mu}\partial_{\mu}A_{\mu}=0$, $A_{\mu}$ being the classical 4-vector potential. This classical condition cannot be satisfied in quantum theory (as the potential components are q-numbers), but it can be expressed by requiring that the allowed states satisfy the eigenvalue equation:

$$
(a_{3}(\vec k)-a_{0}(\vec k))\ket{\psi}=0, \;\forall \; \vec k \;.
$$

Supplementary conditions of this kind, as we shall see, make the states sharp with any number of excitations of the scalar and longitudinal modes unobservable. Hence, as we said, they are traditionally called ghost modes (and their excitations are called ``virtual" particles). 

In combination with this supplementary condition, one can adopt some measures to render the theory mathematically viable despite the presence of a negative norm. One possible approach is the Gupta-Bleuler construction. It resorts to a different metric, \cite{COH}, which we shall call $M$-metric. In this metric, the creation operators of the scalar photons are defined by a modified ``M-adjoint'' operation, with the property that: $a_{0}(\vec k)^{T}\doteq  M a_{0}(\vec k)^{\dagger}M$, where $[a_{0}(\vec k), a_{0}(\vec k)^{\dagger}]=1$ are standard creation and annihilation operators for a harmonic oscillator and $M$ is a unitary, self-adjoint operator, with the property that $M a_{0}(\vec k)^{\dagger}M= - a_{0}(\vec k)^{\dagger}$. It can also be represented as the unitary parity phase operator $(-1)^{a_{0}(\vec k)^{\dagger}a_{0}(\vec k)}$. The new inner product of two general states in this ``M-metric" is then defined as: $$\langle\langle\psi|\phi\rangle\rangle\doteq \bra{\psi}M\ket{\phi}\;.$$ With this new inner product one can check that the $M$-norm of odd-numbered states is positive, as expected. We shall adopt the $M$-metric from now on. 

The interaction Hamiltonian $H_I$ can be written as:
\begin{equation}
H_I= \sum_{\mu}\int {\rm d}^3r\hat j_{\mu}\hat A^{\mu}\;,
\end{equation}
where in our case $\hat  j_{\mu}= (\hat \rho ({\vec r_{ij}}), 0, 0, 0)$ is the source four-vector and $\hat A^{\mu}$ is the potential four-vector operator, with $A^0$ being the scalar potential.  By substituting the expression for the quantised scalar potential in momentum space and for the charge distribution $\hat \rho(\vec r_{ij})$, we have:

\begin{equation}
H_I= \int {\rm d}^3 k qc\sqrt{\frac{\hbar}{2\epsilon_0\omega(k)(2\pi)^3}}[a_{0}(\vec k) \left (b_i^{\dagger}b_i\exp{(-i\vec k \cdot\vec r_i)}+b_j^{\dagger}b_j\exp{(-i\vec k\cdot\vec r_j)}\right )+ {\rm adj.}]\;, \label{HAM}
\end{equation}
where `${\rm adj.}$' denotes the $^T$ operation on the scalar photons sector and the usual adjoint for all other operators (see \cite{COH} for remarks on the negative norm issues when quantizing in the Lorentz gauge, and how this particular definition of adjoint can fix it). 
In the presence of interactions with charges, the supplementary condition defining the Lorenz gauge is modified to:

\begin{equation}
(a_{3}(\vec k)-a_{0}(\vec k)+\hat \eta_{ij}(\vec k))\ket{\psi}=0, \forall \vec k \label{supp}
\end{equation}
where we have defined the operator $\hat \eta_{ij}(\vec k)= c\sqrt{\frac{\hbar}{2\epsilon_0\omega(k)(2\pi)^3}}\hat \rho_{ij}(k)$. 

This condition makes the states sharp with any number of excitations of the scalar and longitudinal modes unobservable. Hence, as we said, they are traditionally called ghost modes (and their excitations are called virtual particles).

We introduce the notation $\hat\eta_{ij}(\vec k)= \frac{g(k)}{\hbar\omega(k)}(b_i^{\dagger}b_i\exp{(-i\vec k \cdot \vec r_i)}+b_j^{\dagger}b_j\exp{(-i\vec k \cdot\vec r_j)})$, where $g(k)=qc\sqrt{\frac{\hbar}{2\epsilon_0\omega(k)(2\pi)^3}}$. The supplementary condition for the Lorenz gauge is expressed by requiring that the allowed states belong to the subspace defined by \eqref{supp}. Since the longitudinal photons are in the vacuum state for this particular problem, we can omit them from the description from now on. Hence the condition \eqref{supp} implies that the allowed states for the scalar photons and the charges are in the span of these states:

\begin{equation}
\ket{\psi}=\ket{c_{ij}}\ket{\lambda_{\vec r_i,\vec r_j}}_0\;.\label{HS}
\end{equation}

Here, the charge sector is described by the state $\ket{c_{ij}}=b_i^{\dagger}b_j^{\dagger}\ket{0}_c$ ($\ket{0}_c$ being the vacuum state of the charge field), which represents the state where one charge (the probe) is located at position $\vec r_i$ and the other (the reference) at position $\vec r_j$; while $\ket{\lambda_{\vec r_i,\vec r_j}}_0$ is an eigenstate of the scalar photon operator $a_{k}$ with eigenvalue $\lambda_{\vec r_i,\vec r_j}= \frac{g(k)}{\hbar\omega(k)}(\exp{(-i\vec k\cdot\vec r_i)}+\exp{(-i\vec k\cdot\vec r_j)})$ (which is the eigenvalue of $\hat \eta_{ij}(\vec k)$ in the state $\ket{c_{ij}}$). The interaction Hamiltonian is responsible for the Coulomb phase (see \cite{MAVE2}for a review of this derivation). Here instead we are interested in the superposition of states satisfying the constraint \eqref{HS}, describing a charge entangled with the field. 

Let us focus on the scalar part of the Hamiltonian only, as it is the only one relevant for this problem (we shall therefore drop the index $0$ in the photon creation and annihilation operator). We can write the scalar Hamiltonian as $H=\int {\rm d}^3 k\hbar \omega(k) (H_{fk} +H_{Ik})$, where

\begin{equation}
H_{fk}=- a(\vec k)^{T}a(\vec k)\;,
\end{equation}

\begin{equation}
H_{Ik}= [a(\vec k) \hat\eta_{ij}(\vec k)^{\dagger}+\hat\eta_{ij}(\vec k) a(\vec k)^{T}] \;.
\end{equation}

Note that the $-$ sign in $H_{fk}$ is the signature of scalar photons: it is a consequence of requiring that the Lagrangian generating this Hamiltonian be Lorentz-covariant and the metric be Minkowski, \cite{HAL, Deser}. 
For simplicity, we consider the situation where a charge $A$ is superposed across two spatial modes, $\vec r_i$ and $\vec r_m$, and another charge $B$ is at a fixed position $\vec r_j$. (The case where the charge B is also superpose is straightforwardly derived from this.)

When there is one charge at $\vec r_i$ and the other at $\vec r_j$, the supplementary condition \eqref{HS} is $\ket{\psi}=\ket{c_{ij}}\ket{\lambda_{\vec r_i,\vec r_j}}_0$;  while the case where the charge $A$ is at location $\vec r_m$, and the other charge at $\vec r_j$, corresponds to the state $\ket{\psi}=\ket{c_{mj}}\ket{\lambda_{\vec r_m,\vec r_j}}_0$. Considering a state where the charge is in a superposition of those two states, e.g. $\ket{\psi_E}=\frac{1}{\sqrt{2}} (\ket{c_{ij}}\ket{\lambda_{\vec r_i,\vec r_j}}_0+\ket{c_{mj}}\ket{\lambda_{\vec r_m,\vec r_j}}_0)$ one can compute the quantity:
\begin{equation}
\langle\langle{\lambda_{mj}}|{\lambda_{ij}}\rangle\rangle = \exp(-\frac{1}{2}|\lambda_{mj}-\lambda_{ij}|^2) =\frac{2g(k)^2}{(\hbar\omega(k))^2 }\left( 1-\cos \left (  \vec k \cdot (\vec r_i-\vec r_m)\right)\right )\;.
\end{equation}

This is the quantity entering the definition of the entanglement between the probe charge and field, as explained in the main text.

\end{document}